\documentclass[structabstract]{aa} 
\usepackage{graphicx}
\newcommand{\ms}{M$_{\odot}$}
\newcommand{\zs}{Z$_{\odot}$}

\newcommand{\mm}{$M_{*}$}
\newcommand{\mex}{$M_{Exp}$}
\newcommand{\mwi}{$M_{Wind}$}
\newcommand{\men}{$M_{Env}$}
\newcommand{\mej}{$M_{Ej}$}
\newcommand{\mpr}{$M_{Proc}$}
\newcommand{\mhe}{$M_{HeC}$}
\newcommand{\mrm}{$M_{Rem}$}
\newcommand{\msa}{$M_{S1}$}
\newcommand{\msb}{$M_{S2}$}

\newcommand{\nn}{$^{14}$N}
\newcommand{\cc}{$^{12}$C}
\newcommand{\oo}{$^{16}$O}
\newcommand{\nea}{$^{20}$Ne}
\newcommand{\neb}{$^{22}$Ne}
\newcommand{\co}{$^{12}$C/$^{16}$O}
\newcommand{\no}{$^{14}$N/$^{16}$O}
\newcommand{\neo}{$^{22}$Ne/$^{20}$Ne}

\begin{document}
   \title{On the origin and composition of Galactic cosmic rays}

   \author{N.Prantzos \inst{1}   }

   \institute{Institut d'Astrophysique de Paris, UMR7095 CNRS, Univ.P. \& M.Curie, 98bis Bd. Arago, 75104 Paris, France;
         \email{prantzos@iap.fr}
             }
   \date{}

 
  \abstract
   {The composition of Galactic Cosmic Rays (GCR) presents strong similarities to the standard (cosmic) composition, but also
   noticeable differences, the most important being the high isotopic ratio of \neo, which is $\sim$5 times higher in GCR than in the Sun.
   This ratio provides key information on the GCR origin.}
    {We investigate the idea that GCR are accelerated by the forward shocks of supernova explosions, as they run
    through the presupernova winds of the massive stars and through the interstellar medium.}
   { We use detailed wind and core yields of rotating and non-rotating models of massive stars with mass loss, as well as simple
   models for the properties of the forward shock and of the circumstellar medium.}
   { We find that the observed GCR \neo \ ratio can be explained if GCR are accelerated  only during the early Sedov
    phase, for shock velocities $>$1500-1900 km/s. The acceleration efficiency is
    found to be of the order of 10$^{-6}$-10$^{-5}$, i.e. a few particles out of a million encountered by the shock
    escape the SN at GCR energies. We also show quantitatively that the widely publicized idea that GCR are accelerated in superbubbles
    fails to account for the high \neo \ ratio in GCR.}
   {}

   \keywords{Cosmic rays - acceleration of particles - abundances  }

   \maketitle
%

\section{Introduction}

Supernova (SN) shocks are generally thought to be the main accelerator of the bulk of
Galactic Cosmic Rays (GCR). Indeed, the power of GCR in the Milky Way is estimated to 
10$^{41}$ ergs/s, corresponding to 10-20 \% of the kinetic power of Galactic
supernovae (assuming canonical values of 2 SN per century, each one releasing 10$^{51}$ ergs
of kinetic energy).

The site of the acceleration of GCR remains debatable today, despite more than five decades
of theoretical and observational studies (e.g. Strong et al. 2007 
and references therein). Over the years, it has been
suggested that GCR are accelerated in 1) SN remnants (either by the forward or the reverse shock
or both), 2) the interstellar medium (ISM), 3) the winds of massive stars, 4) the interiors
of superbubbles, excavated by the massive star winds and the subsequent SN explosions of an OB
association.

Each one of the proposed sites has its own advantages and shortcomings, regarding the
energetics and/or the composition of accelerated matter. For instance, it has been argued that the hot,
low density environment of a superbubble minmizes radiative losses of SN shocks and 
energy losses of accelerated particles, thus allowing the latter to reach substantial energies,
up to the "knee"of the GCR spectrum (e.g. Parizot et al. 2004).
On the other hand, reverse SN shocks running into the SN interior carry insufficient energy
to explain the bulk GCR energetics (Ramaty et al. 1997). Moreover, they should accelerate
$^{59}$Ni, a product of explosive nucleosynthesis which is unstable to e$^-$-capture (with
a lifetime of 10$^4$ yr) and which has not been detected in GCR (Wiedenbeck et al. 1999), while a rapid acceleration
would render it practically stable and thus detectable.

It was realized early on that the {\it elemental} composition of GCR differs significantly
from the one of the ISM. Those differences may provide valuable information
on the origin of GCR particles and, perhaps, on the site - and even the mechanism - of acceleration.
Volatiles behave  differently from refractories: the former display a mass-dependent enrichment
with respect to H, which reaches a factor of 10 for the heaviest of them; the latter are all
overabundant (w.r.t. H) by a factor of 20, while C and O display intermediate overabundances,
by factors of 9 and 5, respectively (e.g. Wiedenbeck 2007 and references therein).

This complex pattern is now thought to result not from ionization effects (as suggested in Cass\'e and
Goret 1978, and further developped by Meyer 1985) but rather from effects related to elemental
condensation temperature (Meyer et al. 1997): refractories are locked in dust grains, which are
sputtered by repeated SN shocks and the released ions are easily picked-up and accelerated (Ellison et al. 1997).
This, quite elaborate, scheme, which builds on earlier ideas by e.g. Cesarsky and Bibring (1981),
accounts quantitatively for most of the observed features of GCR source composition; still, it
leaves unanswered the key issue about the acceleration site of GCR (and how it affects the
composition of accelerated matter).

The most conspicuous feature of GCR source composition   is undoubtely the high isotopic \neo \ ratio.
Its  value was measured since the late 1970ies (Garcia-Munoz, Simpson and Wefel 1979, 
Wiedenbeck and Greiner 1981). The most accurate  measurement today, obtained from
analysis of the CRIS instrument,  leads  to a best estimate  (Binns et al. 2008)
of 0.387 $\pm$ 0.007 (statistical) $\pm$ 0.022 (systematic). 
This is 5.3$\pm$0.3 times the value of the (\neo)$_{\odot}$ \ ratio in the solar wind.
Contrary to the case of the elemental source GCR abundances, which may be affected by various
physico-chemical factors (first ionization potential, condensation temperature, etc.)
isotopic ratios can only be affected by nucleosynthetic processes and thus provide crucial
information on the origin of cosmic ray particles. It should be noticed that up to now there
is now clear evidence for any other GCR isotopic ratio to differ from solar, with the potential exception of
$^{58}$Fe/$^{56}$Fe, which is estimated to 1.5$\pm$0.3 times solar (Binns et al. 2008).

Soon after the discovery of the anomalous GCR \neo \ ratio, Cass\'e and Paul (1982) suggested that
it could be explained by a mixture of $\sim$2\% of material from the wind of a WC star to 98\% of material
with standard composition. In early He-burning, $^{14}$N (produced through the CNO cycle in the previous
H-burning phase) is transformed almost totally in \neb \ through 
$^{14}$N($\alpha,\gamma$)$^{18}$F($\beta^+$)$^{18}$O($\alpha,\gamma$)$^{22}$Ne. He-burning products (like $^{12}$C  and
\neb) are expelled by the stellar winds of  massive stars during their WC phase.  
The observed GCR \neo \ ratio is obtained by assuming dilution of WC material with matter of standard composition.
Subsequent studies put the aforementionned idea on a quantitative basis, with the use of detailed models
of the evolution and nucleosynthesis of massive, mass losing stars (Maeder 1983, Prantzos 1984, Meyer 1985, Prantzos et al. 1987).
In those studies, the acceleration site of GCR was considered as  decoupled from the nucleosynthesis site,
and unrelated to the fraction of admixtured WC material. 

Higdon and Lingenfelter (2003) evaluated quantitatively the \neo \ ratio within a superbubble, created by the
collective action of stellar winds and SN shockwaves. They adopted  stellar wind yields for \nea \ and \neb \ from the models
of Schaller et al. (1992) and SN yields from the models of Woosley and Weaver  (1995). They found that the \neo \ ratio
in the superbubble decreases with time (since \neb \ from the winds dominates the evolution of \neo \ at early times) and that
its time average value is compatible with the GCR source \neo \ inferred from observations. In a subsequent paper, Lingenfelter
and Higdon (2007) recognised that the Schaller et al. (1992) yields of \neb \ were highly overestimated\footnote{The reason 
was the excessively high mass loss rates adopted in that work.} and, consequently, "... new detailed calculations of the
expected GCR isotopic ratio are called for...", but they did not attempt such a re-evaluation. In the meantime,
Binns et al. (2005), using updated wind yields of massive stars with rotation (from the Geneva group, see Sec. 2.2), found
good agreement between the observed \neo \ ratio and an admixture of $\sim$20\% material from WR stars  with 80\% material
of standard composition. According to  Binns et al. (2008), since WR stars are evolutionary products of OB stars, 
such an agreement  "...suggests that OB associations within superbubbles are the likely source of at least a substantial fraction
of GCR".

Howewer,  theoretical studies in the past 10 years  are based mostly on the  paradigm of GCR being accelerated in SN remnants, not in
superbubbles, e.g. \cite{Ptuskin05, Berezhko06,Berezhko09,Ptuskin10,Caprioli10,Schure10,Ellison11} \ 
and references therein. The kinetic energy of the bulk motion of the forward shock of the SN explosion is converted to GCR  energy
through diffusive shock acceleration. The process is highly non-linear and involves the dynamical reaction of both the accelerated
particles and of the magnetic field on the system. Those studies usually take into account the fact that the SN explosion often
occurs within the cavity excavated in the interstellar medium (ISM) by the wind of the massive star prior to the explosion 
(\cite{Biermann01}); however, the structure of the circumstellar environment in that case is quite complex
and simplified models are used for its description. Although \cite{Caprioli11b} considered the composition 
of GCR (H,He, CNO, MgSiAl, Fe)  resulting from such an acceleration site, none of those studies considered the \neo \ ratio.

In this work we study the \neo \ ratio of GCR accelerated by the forward shocks of   SN  explosions, as they run
 through the presupernova winds of  massive stars and through the interstellar medium. We consider the whole mass spectrum 
 of massive stars (from $\sim$10 to 120  \ms), including  stars with either small or large mass losses prior to their
 explosions. We consider stellar properties (masses of winds, ejecta, yields etc.) from recent models with mass loss
 and or without rotation (from Hirschi et al. 2005 and Limongi and Chieffi 2006, respectively), the former 
 having larger \neb \ enhancements in their winds.   
 We adopt a simplified prescription (suggested in Ptuskin and  Zirakashvili 2005 and reformulated in Caprioli 2011) 
 to describe the structure of the  circumstellar medium at the time of the explosion and we consider that GCR start being accelerated 
 in the Sedov-Taylor (ST)  phase of the SN remnant (see e.g. Ptuskin et al. 2010).  By requiring  the resulting 
 IMF averaged \neo \ ratio to equal the observed one  $R_{Obs}$=(\neo)$_{GCR}$/(\neo)$_{\odot}$=5.3$\pm$0.3
 we are able to constrain the forward shock velocity to  values $>$1900 km/s for rotating stars 
 (and to $>$2400 km/s for non rotating ones), i.e.
 we find that GCR are accelerated during the early ST phase, lasting for  a few 10$^ 3$ yr. Assuming that 10\% of the SN
 kinetic energy is converted to GCR, we find that during the acceleration  period a few particles out of a million encountered
 by the forward shock are accelerated.  Finally, we reassess the superbubble paradigm for the origin of GCR, by evaluating
 consistently  the \neo \ ratio with the aforementioned  stellar yields. We find that it can not be as high as observed, unless
 some extremely favorable assumptions are made (only the early period of the superbubble lifetime considered, no gas left over
 from the formation of the OB association). We conclude that superbubbles cannot be at the origin of the bulk of GCR.
 
 The plan of the paper is as follows. In Sec. 2 we present the general "set-up" of our model: the adopted stellar models (Sec. 2.2) 
 and  their wind yields (Sec. 2.3), the description of the circumstellar environment (Sec. 2.4) and the evolution of the forward shock in the
 ST phase (Sec. 2.5). In Sec. 3 we present our results for the (time-dependent) composition of the accelerated particles, the limits
 imposed on the shock velocity by the observed \neo \ ratio and
 the efficiency of the particle acceleration. Finally, in Sec. 4 we explore the \neo \  ratio of GCR,  assumed to be accelerated inside
 a superbubble, and we show that it cannot match the oberved one (unless extreme assumptions are made). The results are
 summarized in Sec. 5.

\section{A toy model for the composition of CR accelerated in massive star winds}

 \subsection{The set-up}
 \label{sub:Toy}

\begin{figure}
   \centering
   \includegraphics[width=8.5cm]{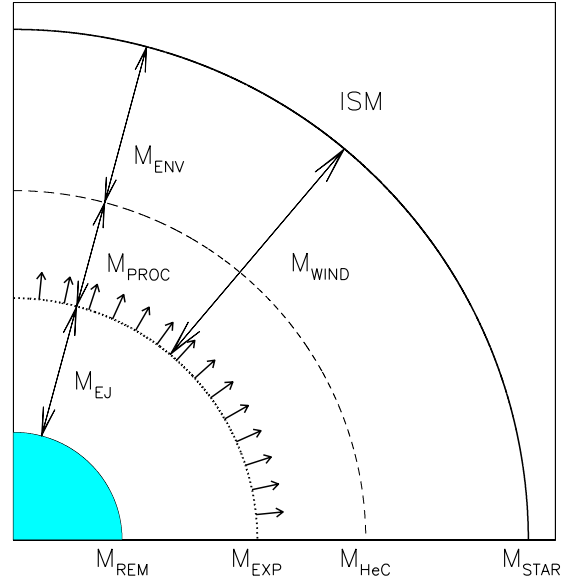}
   \caption{Schematic representation of a supernova exploding in  the wind  of its
   parent star. The star, of initial mass M$_{*}$ explodes with a mass
   M$_{Exp}$, i.e. it has lost a mass M$_{wind}$=M$_{*}$-M$_{Exp}$. The
   most massive stars become WR stars and  their wind expels not only the H-envelope (of mass M$_{Env}$, 
   with composition similar    to the one of the ISM) but also nuclearly processed layers,
   of mass $M_{Proc}$, i.e. M$_{Wind}$=M$_{Env}$+M$_{Proc}$, where M$_{Proc}$=M$_{HeC}$-M$_{Exp}$ and
   M$_{HeC}$ is the mass of the (H-exhausted) He-core. The star leaves a remnant
   (neutron star or black hole) of mass M$_{Rem}$; the mass ejected in the SN explosion
   is M$_{Ej}$=M$_{Exp}$-M$_{Rem}$. Efficient GCR acceleration presumably starts at the beginning of 
   the ST phase, when a mass M$_{S1}\sim$M$_{Ej}$ is swept up
   in front of the SN shock wave (which is indicated by arrows). 
    }
            \label{Fig:SNStruct}
   \end{figure}

The method adopted here in order to calculate the composition of matter accelerated by a single
SN explosion is schematically illustrated in Fig.~\ref{Fig:SNStruct}. At the end of its
life and at the time of its SN explosion, a star of initial mass
\mm \  is left with a mass \mex,   surrounded by a circumstellar shell of mass 
\mwi=\mm-\mex,  which has been lost through stellar wind during its prior hydrostatic evolution
After the SN explosion, a mass of ejecta \mej=\mex-\mrm \ (where \mrm \ is the mass of the compact
remmant, neutron star or black hole) expands first within the shell  of mass \mwi \ and then
in the ISM, with the forward shock having initial velocity $\upsilon_0=\sqrt{2 E_0/M_{Ej}}$,
 where $E_0$ is the kinetic energy of the SN explosion.

In the case of stars ending their lives as WR stars, the wind contains both the original
(nuclearly unprocessed) envelope of mass \men, and nuclearly processed layers of mass \mpr, enriched
in products of H-burning, and in some cases of He-burning as well. For those stars,
\mwi=\mpr+\men \ and \mpr \ is calculated as the difference between the mass of the
nuclearly processed core \mhe \ and the mass at the explosion: \mpr=\mhe-\mex. 
For lower mass stars, exploding as red supergiants, the wind composition results essentially from the 1st dredge-up, i.e. it is a mixture of H-burning products from the
stellar core with the original envelope composition (i.e. mass loss has not uncovered the He-core
at the time of the explosion). The limit between the two classes of stars
depends on their initial mass, mass loss rate and rotational velocity and it is rather poorly 
known at present: in general, in models with  no rotation  stars with \mm$>$32-35 \ms \
become WR stars e.g. (Heger et al. 2002), while in models 
with rotation that limit may be as low as 22 \ms \ (Meynet and Maeder 2000).

The first phase of the supernova remnant ("free expansion") 
takes place at shock velocity $\upsilon\sim const.$ and ends when a mass 
\msa$\sim$\mej \ has been swept up in front of the shock wave, at which point the
ST phase sets in. Following Ptuskin et al. (2010), we assume that efficient GCR acceleration
starts at this time, where the situation is energetically most favorable.
In our baseline model we shall consider constant acceleration efficiency ;
time-dependent efficiency
of particle acceleration is the subject of current researches (see Ellison and Bykov 2011, 
Drury 2011, and references therein) and will be briefly discussed in Sec. 3.3. 

The ST phase proceeds adiabatically, i.e. at $\sim$constant energy and
with decreasing velocity, until
the temperature of the gas engulfed by the shock front  drops to levels allowing a significant fraction
(about 50\%) of the remaining energy to be radiated away. At that time, an amount  
of matter \msb$>>$\mej \ has been swept up and the shock enters the "snow-plow" phase. At this
point - and, perhaps, even earlier, during the ST phase - the forward shock is 
too weak to accelerate particles to GCR energies any more.

In the aforementioned scenario, GCR are accelerated from a pool of particles with composition
characteristic of the mass \mwi \ early on. Depending on the initial stellar mass, this
composition may be rich in products of H- (and He-) burning. It is progressively diluted
with ambient (first wind - with normal \neb - and then interstellar) gas and at the end of the ST
phase it ressembles closely the one of the ISM.  The  GCR source composition observed on 
Earth should correspond to the  average composition 
between the early ST phase and some later evolutionary stage of the remnant, 
and should result from the whole mass spectrum
of exploding stars, i.e. it should be averaged over a stellar initial mass function (IMF).

\subsection{Properties of mass losing stars}
\label{sub:StarProp}

   \begin{figure}
   \centering
   \includegraphics[width=0.495\textwidth]{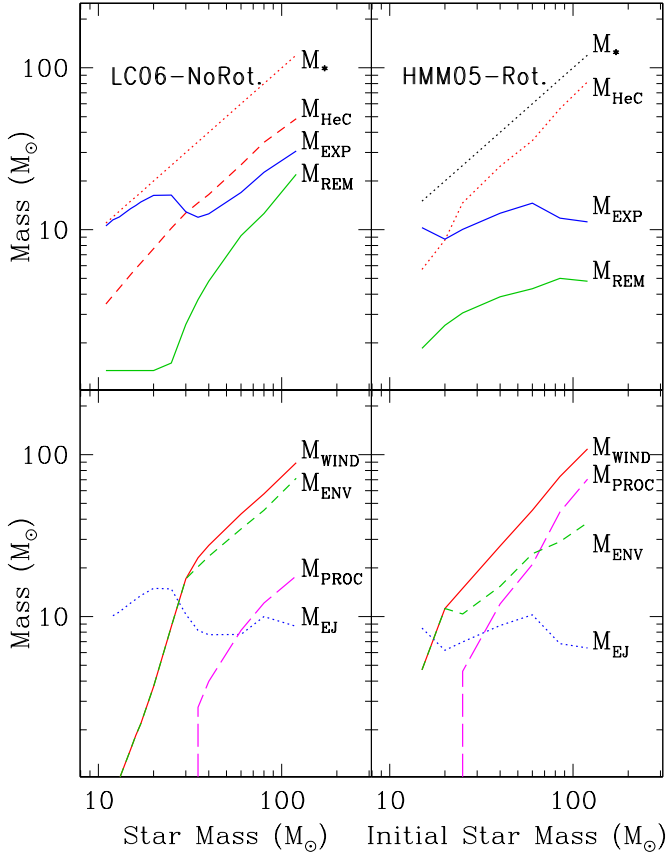}
   \caption{Input data adopted in this work, for solar metallicity massive stars with mass loss.{\it Left}:
   non rotating models from Limongi and Chieffi (2006); {\it Right:} Rotating models from Hirschi, Meynet and Maeder (2005).
    {\it Top:} Mass of the (H-exhausted) He core
   $M_{HeC}$, of the star at explosion $M_{Exp}$ and of the compact remnant $M_{Rem}$ as a function 
    of the initial stellar mass $M_{\bf *}$. {\it Bottom:}
   derived quantities are the mass of the wind $M_{Wind}$=$M_{\bf *}$-$M_{Exp}$, the mass of the
   nuclearly processed layer $M_{Proc}$=$M_{HeC}$-$M_{Rem}$ and of the unprocessed (\neb-normal) envelope
   $M_{Env}$=$M_{Wind}$-$M_{Proc}$ and the mass of the SN ejecta $M_{Ej}$=$M_{Exp}$-$M_{Rem}$
   In all cases, curves are smoothly interpolated between model results.
    }
\label{Fig:SNmasses}
   \end{figure}

We adopt two sets of stellar models in this work. They are calculated
for stars of solar metallicity, and in both cases  the solar mixture of Anders and Grevesse
(1979) is adopted. The corresponding metallicity is \zs=0.019, substantially larger than
more recent values (Lodders 2003, Asplund et al. 2010) and this difference results
in particular from the reduction in the past decade of the solar abundances of C, N, O and Ne, which
are key elements for the purpose of this work. For obvious consistency reasons, we keep here the
Anders and Grevesse (1979) values, when comparing  our results for GCR to solar ones.

The first set of stellar models 
is the one of the Frascati group (Limongi and Chieffi 2006, hereafter LC06). It concerns 15 model 
stars between 11 and 120 \ms \  with mass loss but no rotation. The model includes
all stages of hydrostatic nuclear burning
and simulates the final stellar explosion by imparting an initial velocity 
 to a mass coordinate of 1 \ms \ (i.e. well inside the Fe core of the stars); 
the  mass cut (the limit separating the ejecta of mass \mej \ from the compact remnant of mass \mrm) is chosen such as  0.1 \ms \ of $^{56}$Ni is ejected by the explosion.

The various masses involved in the "toy model" of Sec.~\ref{sub:Toy} and Fig.~\ref{Fig:SNStruct}
are provided in Table 2 of LC06 and are displayed in Fig.~\ref{Fig:SNmasses} ({\it top left}) of this work, 
whereas derived quantities are displayed in Fig.~\ref{Fig:SNmasses} ({\it bottom left}).
It can be seen that stars with \mm$<$20 \ms \ have lost a negligible
amount of mass prior to the explosion (\mwi$<$\mej) and the ST phase starts within the
ambient ISM. 
Stars with \mm$>$30 \ms \ have \mhe$>$\mex, i.e. they explode as WR stars, having
expelled nuclearly processed layers in their winds. But only for the most massive stars
(\mm$>$60 \ms) one has \mej$<$\mpr, i.e. in the beginning of the ST phase the shock wave still
expands into material with composition reflecting the one of the nuclearly processed core; for lower
mass stars (30$<$\mm/\ms$<$60) the ST phase starts when the shock wave encounters material with mixed composition from the core and the envelope, i.e. less enriched in H- and He- burning products.
An interesting feature of those models is that the mass of the ejecta \mej$\sim$10 \ms, is similar
for all of them, leading to similar properties(duration, swept-up mass)
 for their corresponding ST phases.

The second set of stellar models is the one of the Geneva group, calculated by 
Hirschi, Meynet and Maeder (2005, hereafter HMM05) and includes both mass loss and rotation for stars with mass between 12 and 60 \ms; it has been complemented with results for stars of 85 and 120 \ms, kindly provided by G. Meynet (private communication). The initial rotational velocity is
$v_{Rot}$=300 km/s on the ZAMS, corrrsponding to an average velocity of 220 km/s on the main sequence,
i.e. close to the average observed value. Non-rotating models with the same physical ingredients (for convection, mass loss, etc.)
have also been calculated, for comparison.
The evolution has been calculated to the end of Si-burning but the final SN explosion was not considered. Instead, an empirical prescription was used to evaluate the mass of the compact remnant.
The masses of our "toy model", as given in Table 2 of HMM05, are displayed in Fig.~\ref{Fig:SNmasses}.
There is an important difference with respect to Table 2 of HMM05: they provide the mass
of He-core at the time of the explosion and for the most massive stars this coincides
with \mex \ (i.e. the mass left to the star at explosion 
is smaller than the maximum extent of the He-core); however, we are interested at the true value of 
\mhe \ (since this will determine how much mass of processed material the shock wave will encounter) and
that value is obtained through the detailed results of HMM05 (displaying the wind composition
as a function of time - or of mass left).

Comparing the results of LC06 and HMM05 one sees that rotation increases the mass loss (\mwi \
larger in HMM05), thus leaving the star with a smaller mass at explosion (\mex \ smaller in HMM05).
Rotation also increases the size of nuclearly processed regions (\mhe \ larger in HMM05), since 
matter is rotationally mixed outwards to larger distances
than achieved through convection. In turn, this leads to  larger amounts of processed material
\mpr \ for the HMM05 models.

The aforementioned features of rotating vs non-rotating models, which are explained in detail in e.g.
Maeder and Meynet (2000) are crucial in understanding the differences in the corresponding 
wind yields of the stars.

\subsection{The wind composition of massive stars}

LC06 provided (private communication)  {\it yields} $y_i$ of all stable nuclear species, from H to Ge,
included in their models and ejected through the winds of the stars, up to the moment of the explosion.
HMM05 provide (Table 3 in their paper) the {\it net yields} $y_{n,i}$ of the winds of
their models for stable species from
$^3$He to $^{23}$Na, from which the yields can be recovered through
\begin{equation}
y_i(M_*) = y_{n,i}(M_*) + {M_{Wind}} {\rm X_{\odot,i}}
\end{equation}
where the adopted solar values X$_{\odot,i}$ are displayed in Table 1 of HMM05.

\begin{figure}
   \centering
   \includegraphics[width=0.49\textwidth]{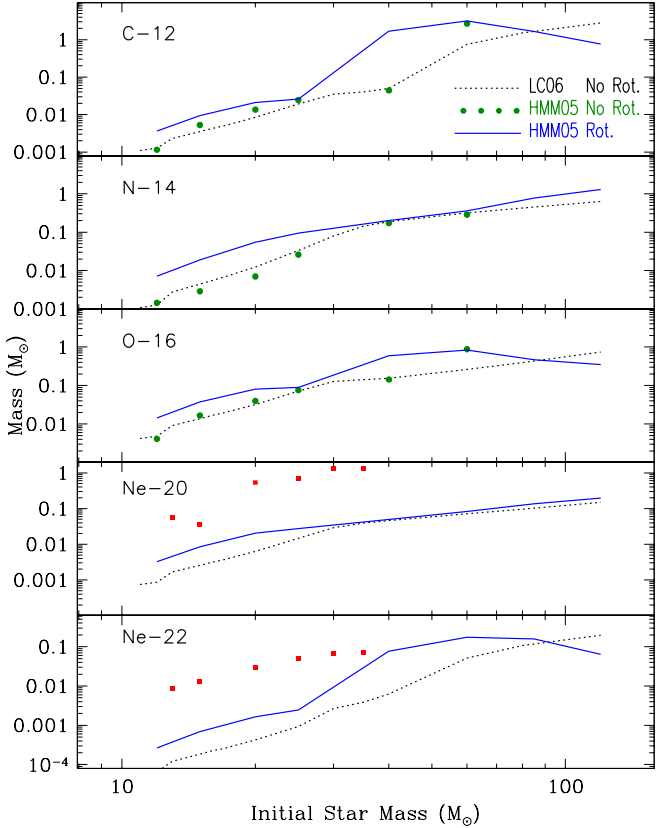}
   \caption{Total masses of selected isotopes in the winds of the massive stars, from the
   two sets of models adopted in this work. {Solid curves}  are yields from HMM05 (with rotation)
     and {dotted curves}  from LC06 (no rotation),
      in both cases interpolated between model results.
    The points correspond to results of the HMM05 models with no rotation and,  in general,
     agree well with LC06 results except for the 60 \ms \ star.
   }
         \label{Fig:windmasses}
   \end{figure}

The wind yields of a few selected species appear in Fig.~\ref{Fig:windmasses}, for 
 the non-rotating models of LC06 and
for both the non-rotating and the rotating models of HMM05. It can be seen that, in general,
there is excellent agreement between the results for non-rotating models  of LC06 and HMM05,  for stars
up to 40 \ms. Their results differ only for the 60 \ms \ model (and presumably for higher masses
as well) and only for the cases of the He-burning products $^{12}$C, $^{16}$O and $^{22}$Ne.
Since both HMM05 and LC06 use the same prescriptions for mass loss, the reason of that discrepancy could
be the use of a small amount of overshooting in the case of HMM05.

Rotation has a twofold effect on stellar yields: it increases the size of the nuclearly processed layers
(since it mixes material further than convection alone) and reduces the escape velocity in the stellar equator,
allowing larger amounts of mass to be ejected in the wind. Both effects enhance the wind yields up to some
mass limit; above it, the wind has removed so much mass, that less material is left in the star 
to be processed in subsequent stages of the evolution, thus reducing the corresponding yields. This
is the case, for instance, with the He-burning products \cc \ and \neb, the yields of which decrease above
$\sim$60 \ms \ in the rotating HMM05 models (see Fig. \ref{Fig:windmasses} and HMM05 for details).

In the following we assume  that the wind interaction
with the ISM has not substantially changed the wind stratification:
 the forward shock will first encounter
the innermost wind layers, containing processed material in the case of  the most massive stars;  
later it will encounter the outer wind layers (containing mostly the initial composition), before
running into the ISM.  Fig. 4  displays  the {\it mass integrated composition of the wind} (for a few key metals),
 as encountered
by the forward shock, moving outwards from \mex, in the case of  two rotating model stars of
 25 \ms \ and 60 \ms \ (from HMM05). The quantity
 \begin{equation}
 m_i(M) \ = \ \int_{M_{Exp}}^{M_*}X_{Wind,i}(M) \ dM
 \end{equation} 
is displayed as a function of mass coordinate $M$, X$_{wind,i}(M)$ being the mass fraction of isotope $i$
in the wind of star of mass \mm. Obviously, one has $m_i$(\mex)=0. and  $m_i$(\mm)=$y_i$(\mm), i.e.
at crossing the last (outermost) wind layer, the forward shock has encountered the totality of the yield $y_i$(\mm).

\begin{figure}
   \centering
 \includegraphics[angle=-90,width=0.49\textwidth]{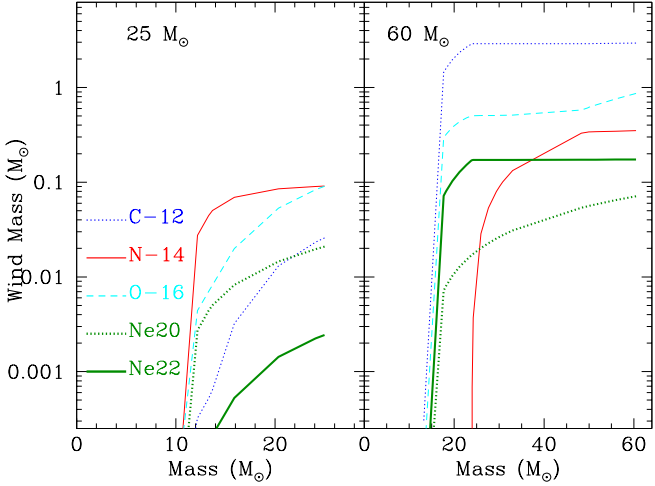}
 \caption{ Composition encountered by the forward shock wave as it moves
 (from left to right) through the winds of 25 \ms \ (left) and 60 \ms (right) rotating stars.
 The shock wave starts at \mex =10 \ms and 14.8 \ms, respectively, and the ST phase
 sets in when a mass M$_{S1}$=\mex \ has been swept up and, presumably, well mixed.  
 The quantities displayed correspond to the mass integrated composition of the wind,
 starting at \mex. 
 }
         \label{Fig:wind_Struct}
   \end{figure}

An inspection of Fig.~\ref{Fig:wind_Struct} shows that:

- in the case of the 25 \ms \ star (no He-burning products encountered by the shock wave),
 the innermost layers contain most of the \nn \ produced by the CNO cycle (its mass increases
 slowly in the outermost layers), and little of \cc \ and \oo, which are depleted by the CNO cycle
  (their mass increases rapidly in the outermost layers); \nea \ and \neb \ are little affected by H-burning
  and their integrated wind mass increases in a way intermediate between the  cases of \nn \ and \cc.

- in the case of the 60 \ms \ star (He-burning, then H-burning products encountered by the shock wave),
the quasi-totality of the \cc\ and \neb \ produced by He-burning are encountered in the inner layers
(inside $\sim$25 \ms), while essentially no \nn \ is left in that region; the majority of \nn \ is found in the
region 25$<M$/\ms$<$50  while substantial  \oo \ is found in the unprocessed envelope (beyond 50 \ms).
Finally, \nea \ is little affected everywhere ($\sim$const. mass fraction) and its integrated mass rises regularly with
the wind mass of the star.

If detailed information (i.e. wind composition as function of wind mass coordinate) is available,
the composition of the material swept-up by the forward shock can be calculated in a consistent way. In case only the
wind yields $y_i$ and the various characteristic masses of Fig. 2 are available, one may adopt the following approximations:

(i)  Stars never reaching the WR stage (\mm$<$22 \ms \ for rotating models and \mm$<$34 \ms for non rotating ones)
have their envelopes fully mixed after the 1st dredge-up episode in the red supergiant phase; the wind composition
is then simply obtained as:
\begin{equation}
X_{wind,i}(M_*)  \ = \  {{y_i(M_*)}\over{M_{Wind}(M_*)}}
\end{equation}

(ii) Stars reaching the WR stage eject first their unprocessed envelope of mass \men \ and
composition X$_{env,i}$=X$_{ISM,i}$
and then the innermost processed layers of mass \mpr \ and composition X$_{PROC,i}$ such
that
\begin{equation}
y_i(M_*) \ = \ X_{Env,i} M_{env}(M_*) \ + \  X_{Proc,i} M_{Proc}(M_*)
\end{equation}
which allows one to derive  X$_{Proc,i}$, i.e. an average abundance in the processed layers, and obtain thus
an average wind composition profile

\begin{equation}
X_{Wind,i}(M)  \ = \  X_{Proc,i}(M_*)  \ \ {for } M_{Exp}<M< M_{HeC} \\
\end{equation}
\begin{equation}
X_{Wind,i}(M)  \ = \  X_{Env,i}(M_*) \ \ {\rm  for} M_{HeC}<M< M_*
\end{equation}


The method outlined here turns out to provide a good approximation for the detailed composition 
profiles of the HMM05 models and has been used here for both sets of LC06 and HMM05 models,
for consistency reasons. Taking ito account all the uncertainties of the models (prescriptions for mass loss,
convective and rotational mixing etc.),  we consider the aforementioned approximation
as fairly satisfactory for the purpose of this work.

\subsection{The circumstellar environment of massive stars}

According to the ideas outlined in Sec. 2.1 particle acceleration to GCR energies
starts when a mass $M_{S1}\sim$\mej \ has been swept-up by the forward shock.
The composition of that material, at the very beginning of particle acceleration,
is the one of the material located at mass coordinate $A_1$=\mex + \mej \
and constitutes obviously an extreme for the abundances of the corresponding elements
(upper limit for those produced in the stellar interior and lower limit for those destroyed, like H).
In the following we shall assume that at the time of the explosion  the stellar wind 
has fully kept the stratification of its various layers, as they were progressively leaving the
stellar surface. 

 The circumstellar region hit by the forward shock has a complex structure, depending on the
 properties of the exploded star. That structure has been explored in some detail with hydrodynamical models
 by Garcia-Segura et al. (1996a,b) for non-rotating stars. In the case study of a 35 \ms \ star, they find
 that the rapid wind of the O star excavates a large bubble (radius $\sim$36 pc) of low density (10$^{-3}$ cm$^{-3}$).
 Inside it propagates a dense ($\sim$1 cm$^{-3}$), slow wind - released during the red supergiant
phase - which occupies the innermost few pc (depending on its assumed velocity, of the order of a few tens of km/s).
The subsequent fast ($\sim$10$^3$ km/s) WR wind compresses the RSG wind, and most of the mass of the latter is
found within a thin shell.
 
 The aforementioned study illustrates the complexity of the situation, but its results can hardly be
 generalized to the whole mass range of
 massive stars (for instance, lower mass stars will not display the fast WR wind). 
 In fact, its results cannot even be safely used for the 35 \ms \ star, since they
 depend so critically on the adopted parameters of the model (mass loss rates and wind velocities for the
 various stages). And they certainly fail to describe the situation for rotating stars, which display
 slow but intense (and not radially symmetric) mass losses on the Main sequence.
 
In view of these uncertainties, we adopt here a simplified prescription for the
structure of the circumstellar bubble, assuming spherical symmetry
in all cases. We assume that the winds have excavated a bubble of {\it mean density}
$n_0$=0.01 cm$^{-3}$ and, consequently of radius 
 \begin{equation}
 R_B \ = \  \left(\frac{3 \ M_W}{4 \pi \ \rho_0}\right)^{1/3}
 \end{equation}
 with $\rho_0$=$n_0 \ m_p$,  $m_p$ being the proton mass.
Inside the bubble, the density profile is $\rho(r) \propto r^{-2}$, i.e. it corresponds to a
 steady stellar wind with mass loss rate ${\dot M_W}$ and velocity $v_W$, which is given by
  \begin{equation}
 \rho(r) \ = \  \frac{\dot M_W}{4 \pi v_W r^2}
 \end{equation}
Our choice of $\rho_0$ automatically fixes the  $\rho(r) $ profile and corresponds to a combination
of ${\dot M_W}$ and $v_W$ values. Obviously, one has
\begin{equation}
M_W \ = \ \int_0^{R_W} \ \rho(r) \ dr
\end{equation}
Outside $R_W$ we assume an ISM with constant density
$\rho_{ISM}$=1 $m_p$ cm$^{-3}$.
Our approach is similar to Caprioli (2011), but we do not consider here
the more complicate case of a WR wind overtaking a RSG wind.

\subsection{Evolution during the Sedov-Taylor (ST) phase}

We follow the propagation of the forward shock first through the wind bubble and then through the ISM
 with a simple model presented in Ptuskin and Zirakashvili (2005) and, in a more concise form, in
 Caprioli (2011, his eqs.
   3.4 to 3.9).  We start the calculation from the "free-expansion"(ejecta dominated) phase, where
 the swept-up mass is smaller than \mej \ and which can be described by self-similar
 analytical solutions. In the subsequent ST  phase (swept up mass $<$ \mej),  the model is based on the 
 "thin shell"approximation (e.g. Ostriker and McKee 1988)
 which assumes that the swept up mass is concentrated in a thin shell behind the shock.
 We solve numerically the time-dependent equations
 for continuity of mass, energy and momentum, to recover shock radius and velocity as a function of time. 
 Unlike Caprioli (2011) we assume full adiabaticity
 in the ST phase, i.e. we do not take into account the 10-20\% energy losses of the shock 
 through CR acceleration, which would reduce the shock velocity ($\propto E_0^{0.5}$) by less than 10\%.
 The mass swept up in the ST phase inside shock radius $R_S$ is
 \begin{equation}
 M_S(<R_S) \ = \  M_{EJ} \ + \ 4  \pi \int_{R_{S1}}^{R_{S}} \rho(r) r^2  dr
 \end{equation}
We follow the evolution all the way through the  ST  phase, which  
ends when a significant fraction of the energy of the cooling remnant is radiated
away (through recombination emission); for a solar mixture this occurs at time
\begin{equation}
t_{S2} \ \sim 4.4 \ 10^4 \ {\rm yr} \  \left({{E_0}\over{10^{51}{\rm erg}}}\right)^{2/9}
\left({{n_{ISM}}\over{{\rm cm^3}}}\right)^{-5/9} 
\end{equation}

In the framework of this simple model we are able to calculate the composition 
of the material encountered (and presumably accelerated)
by the forward shock as a function of time, or of the swept-up mass: 
indeed, the integrated mass of each element swept up by the 
forward shock is given by an equation similar to Eq. 2 
\begin{equation}
 m_i(M_S) \ = \ \int_{M_{exp}}^{M_S(<R_S)} X_{i}(M_S) \ dM_S
 \end{equation} 
 where $X_{i}(M_S)$=X$_{wind,i}$ for $M_S<$\mm \ (Sec. 2.3) 
 and $X_{i}(M)$=X$_{ISM,i}$ for $M_S>$\mm , i.e. when
 the shock propagates in the ISM; the upper limit in the integral is
 given by eq. (10).
 
Eq. 12 allows one to link the stellar model, i.e. the abundance profiles $X(M)$, to the evolution during the ST
phase  through Eq. 10,  and to the properties of the shock wave.
 Since the mass M$_{S2}$ swept-up in the end of the ST phase is much larger than 
 the wind mass in all cases (a few 10$^3$ \ms, compared to $\sim$100 \ms at most)
for the largest part of the ST phase the swept up material has ISM composition. In order
to obtain significant deviations from the   solar composition, such as the observed \neo \ ratio, one should assume that 
{\it significant acceleration  occurs  only in the early ST phase}, when the
forward shock is stronger and its velocity higher.
 
 \section{Results}
 
\subsection{Composition of matter in the ST phase}

   \begin{figure}
   \centering
 \includegraphics[width=0.49\textwidth]{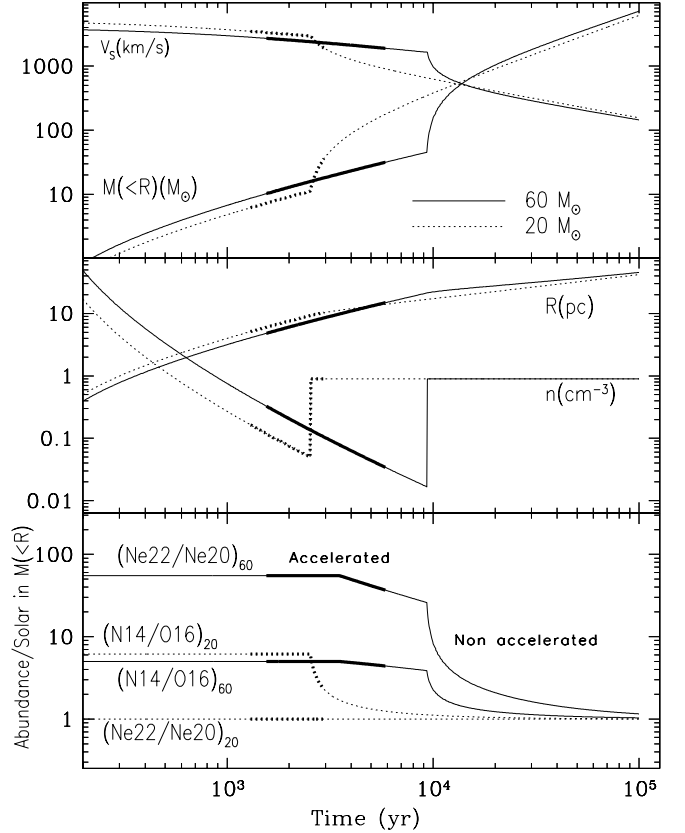}
 \caption{ Evolution of the SN remnants of two rotating stars of initial masses
 20 \ms \ ({\it dotted}) and  60 \ms \ ({\it solid}), respectively. {\it Top}: 
 velocity of the forward shock 
 and  swept up mass. {\it Middle}: Shock radius and density profile before shock arrival
 (see text).
 {\it Bottom:} Composition of the matter swept up up to time $t$ 
 by the shock wave,
 for the \no \ and \neo \ ratios. In all panels, the {\it thick } portions of the curves indicate
 the period of efficient particle acceleration, i.e. from the beginning of the ST phase and as far as 
  $\upsilon_s>\upsilon_{min}$.
 The value of $\upsilon_{min}$ (=1900 km/s for the rotating star models in the figure) is chosen 
  so that the IMF averaged theoretical ratio  of \neo \ 
 matches the observed one in GCR (see text, Eq. (13)  and Fig.~\ref{Fig:ModelsVsObs}). 
 }
         \label{Fig:GCRaccel}
   \end{figure}

Fig. \ref{Fig:GCRaccel} (top) displays the evolution of the velocity $\upsilon_S$ 
and radius $R_S$ of the forward shock and of the mass $M_S(<R_S)$ swept up
by it, for the cases of a 20 and a 60 \ms \ rotating star, respectively.
 The density of the unperturbed ISM is taken to be 1 cm$^{-3}$ in all cases.

The similarity of the curves for  $\upsilon_S$, $R_S$ and $M_S(<R_S)$ for the 20 and
60 \ms \ stars simply reflects the self-similarity of the ST solution. The small differences
in the early ST phase are due to the difference of the ejected mass \mej \ in the two stars
(6 vs 10 \ms, see Fig. 2), since the other parameters ($E_0$ and $n_{ISM}$) are the same.

The bottom panel of Fig.  \ref{Fig:GCRaccel} displays the evolution of the \no \ and  \neo  \
ratios (i.e. the ratios of the corresponding masses of Eq. 12  for each isotope), 
 expressed in units of the corresponding solar values. In both cases, the forward shock
first encounters layers with low \oo \ and high \nn, resulting in a high \no \ ratio (in the 60 \ms \ star,
\nn \ is depleted from He-burning in the innermost layers, resulting in a slightly smaller
\no \ ratio than in the 20 \ms \ case).
Subsequently, in the 20 \ms \ star the shock moves rapidly through the small remaining 
stellar envelope (3.4 \ms) and starts propagating in the ISM, thus decreasing  rapidly its
mass-integrated \no \ ratio. 
In the 60 \ms \ star, the shock runs through $\sim$20 \ms \ of processed material, with
a high value of \no, before getting to the ISM; the corresponding \no \ ratio decreases
more slowly than in the 20 \ms \ case.

The evolution of \neo \ is quite different in the two models. In the 20 \ms \ star,
no He-burning products are encountered by the shock wave and the \neo \ ratio
has always its initial (solar) value. A high \neo \ ratio is initially 
encountered in the processed layers
of the 60 \ms \ star, which is progressively diluted as the shock moves outwards.

   \begin{figure}
   \centering
 \includegraphics[width=0.49\textwidth]{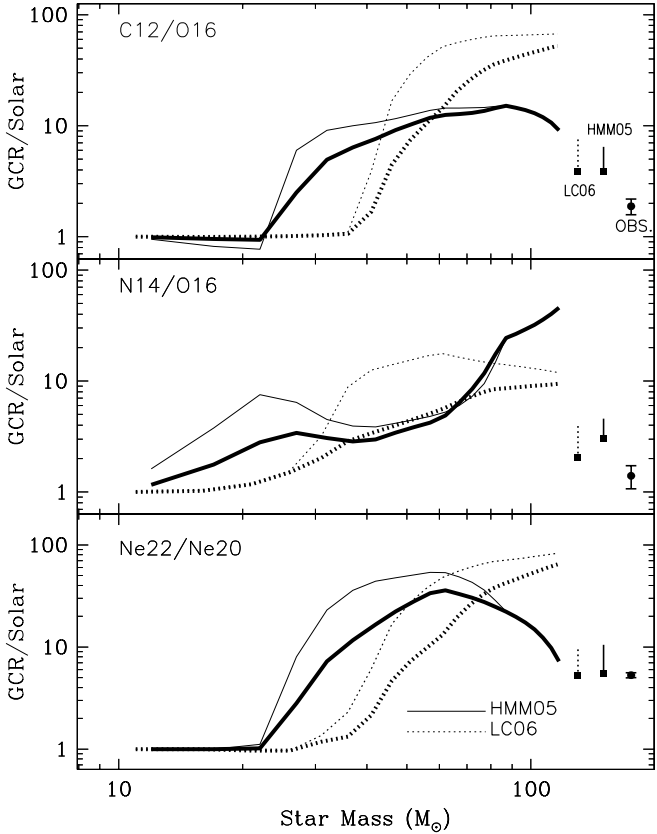}
 \caption{Abundance ratios of various nuclear species in GCR source normalized to 
 the corresponding solar ones,  as a function of the
 initial stellar mass. In all panels, {\it solid curves} correspond to models of
  HMM05 and {\it dotted curves} to models of LC06. {\it Upper,  thin curves} 
  are for  GCR  accelerated at the beginning of the ST phase 
  and {\it lower,  thick curves } for the time-average at the end of GCR acceleration. 
  An average over a Salpeter IMF (and accounting
 for the swept up mass in each case) produces the vertical segments to the right, their top
 point corresponding to the beginning and the bottom one to the end of the GCR acceleration
 phase, respectively (also indicated by {\it filled squares}). These results are 
 compared to GCR source abundance
 ratios as derived by ACE data (points at the extreme  right 
 with error bars) in Binns et al. (2005). The most significant,
 unaffected by FIP, volatility etc., is the one of $^{22}$Ne/$^{20}$Ne. The end of the GCR
 acceleration phase is assumed to correspond to shock velocities $\upsilon_{min}$ such that
the time and IMF averaged theoretical ratio (squares) of $^{22}$Ne/$^{20}$Ne
 matches the observed one (see text). For the
 set up adopted here we find we find $\upsilon_{min}$=1900 km/s for HMM models and 
 $\upsilon_{min}$=2400 km/s for LC06 models.}
         \label{Fig:ModelsVsObs}
   \end{figure}

\subsection{Composition of accelerated particles}

The composition curves of the bottom panel of Fig.  \ref{Fig:GCRaccel} give the time
(or mass) integrated composition encountered by the shock wave and, in consequence,
the composition of the particles that have been accelerated {\it up to that time}.
We assume here that {\it particles are accelerated to GCR energies with the same efficiency  
for shock velocities
higher than some critical value $\upsilon_{min}$, which is the same for all stellar masses}.
We determine $\upsilon_{min}$ empirically by requiring that, when averaged over a stellar Initial mass function
(IMF) $\Phi(M_*)$,  the ratio
\begin{equation}
\frac{M_{22}}{M_{20}} = \frac{\int_{10 M_{\odot}}^{120 M_{\odot}} dM_* m_{22}(M_*) \Phi(M_*)}{\int_{10 M_{\odot}}^{120 M_{\odot}} dM_* m_{20}(M_*) \Phi(M_*)} =
R_{Obs} \frac{X_{22,\odot}}{X_{20,\odot}}
\end{equation}
where $R_{Obs}$=5.3$\pm$0.3  is the observationally determined source GCR ratio of \neo \ in solar units and $m_{22}(M_*)$ and $m_{20}(M_*)$
are calculated from Eq. 12 for stars of mass \mm \ and for swept up masses $M_S(\upsilon>\upsilon_{min}$).

We adopt here a Salpeter IMF $\Phi(M_*) \propto M_*^{-X}$ with $X$=2.35.
The results of the procedure appear in Fig. \ref{Fig:ModelsVsObs} for a few selectedabundance ratios and for the models
of both HMM05 ({\it solid curves}) and LC06 ({\it dotted curves}). In all panels, the upper ({\it thin}) curves
correspond to the composition accelerated at the beginning of the ST phase (maximal possible deviations
from solar composition).  It can be seen that stars with mass $<$22 \ms \ (for HMM05) and 32 \ms \
(for LC06) display no He-burning products in their accelerated particles. 
N is overabundant in lower stellar masses, due to the 1st dredge-up.

The lower ({\it thick}) curves  in all panels of Fig. 6 correspond to composition accelerated {\it up to the end of the
acceleration period} which is assumed to occur for a shock velocity $\upsilon_{min}$.
The corresponding IMF-averaged quantities (between 10 and 120 \ms) 
are displayed on the right of the curves: their uppermost point
corresponds to the beginning of the ST phase and the lower one (also indicated with a {\it filled square})
to the end of the acceleration period, i.e. to  $\upsilon_{min}$. The value of $\upsilon_{min}$
is found to be $\sim$ 1900 km/s  for the HMM05 models with rotation and $\sim$20\% higher (2400 km/s)
for the LC06 models without rotation. The reason for that difference is, of course the fact that rotating models
have larger processed layers, requiring  more dilution with circumstellar material. 

The top and middle panels of Fig. 6 display the corresponding ratios for \co \ and \no, respectively.
In both cases, the IMF averaged ratios are higher than the observed ones in GCR  (which are, in their turn,
higher than solar), by factors of 1.5-2. Unlike \neo, these are ratios of different elements, having
different atomic properties. GCR source abundances are known to be affected by atomic effects, e.g. First
Ionization Potential (FIP), or, perhaps more plausibly, volatility and mass/charge ratio
(see extensive discussion in Meyer et al. 1997). The analysis of such effects is beyond the scope
of this study. We simply notice here that observations indicate that
refractory elements are relatively more abundant in 
GCR sources than volatiles. Meyer et al. (1997) attribute that to the fact that refractories are
locked up in grains, which are sputtered by the shock wave and the released ions are easily picked up
and accelerated. \oo \ being more refractory than both \nn \ and \cc, it is expected that its abundance
in GCR source will be enhanced by that effect, and the corresponding \co \ and \no \ ratios in GCR source 
will be lower than predicted from stellar nucleosynthesis alone. The fact that in Fig. 6 we obtain
higher than observed ratios for \co \ and \no \ is encouraging in that respect, since the aforementioned atomic
effects would lower those ratios, hopefully to their observed values\footnote{One might think that a combined
analysis of the \co \ and \no \ ratios of Fig. 6 could help to constrain the atomic processes shaping the GCR source 
abundances. However, the abundances of \cc \ and \oo \ are affected by the still uncertain  value
of the \cc($\alpha,\gamma$\oo \ reaction rate and are unsuitable for such a study.}.

  \begin{figure}
   \centering
 \includegraphics[width=0.49\textwidth]{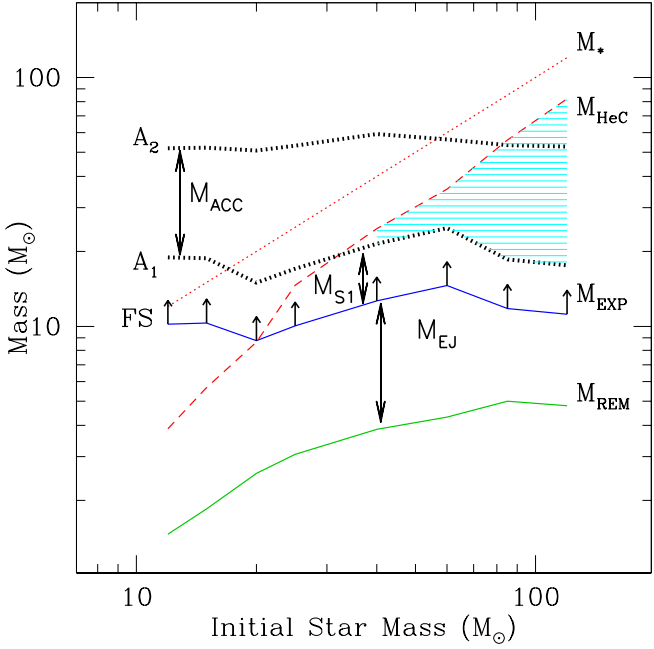}
 \caption{Graphical presentation of the results discussed in Sec. 3.3. Particle acceleration
 starts at the beginning of the ST phase, located at mass coordinate $A_1$=\mex + \mej, i.e. when
 the forward shock (FS, {\it arrows}), launched at \mex,  has swept up a mass $M_{S1}$=\mej. Acceleration stops
 at mass coordinate $A_2$, corresponding in the case dicussed here to a shock velocity of 1900 km/s.
 The mass sampled by the FS between those two regions is $M_{Acc}$=$A_2$-$A_1$. For rotating stars with
 mass $M>$30 \ms, an increasing part of $M_{Acc}$ includes nuclearly processed material ({\it shaded aerea}),
 while for rotating stars with $M<$18 \ms, $M_{ACcc}$ contains only material of ISM (=solar) somposition.
 }
         \label{Fig:AccelvsM}
   \end{figure}

 \subsection{Efficiency of particle acceleration}
 
 The material of Sec. 3.1 and 3.2 is summarized in Fig. \ref{Fig:AccelvsM} for the rotating models of HMM05. The forward
 shock, launched at mass coordinate \mex, sweeps up a mass $M_{S1}\sim$\mex \ and then
 starts accelerating particles, at mass coordinate $A_1$=\mex+$M_{S1}$, up to point $A_2$ (where
 its velocity becomes $\upsilon_{min}$).
 For rotating stars with $M<$15 \ms, $A_1$ lies beyond the stellar surface and only ISM is accelerated.
 In stars with 15$<$M$_*$/\ms$<$25, $A_1$ lies beyond the processed/mixed 
 interior \mhe \ and the forward shock accelerates first envelope and 
 then ISM material. For stars above 35 \ms, the shock first
 accelerates processed material (hatched aerea), then the - \neb \ normal - envelope and then ISM. Finally, for stars with M$>$70 \ms, particle acceleration ends when the shock is still within the massive stellar envelope.
   
   The mass of circumstellar material
from which particles are accelerated is  $M_{ACC}=A_2 - A_1$ and lies in the range 30-40 \ms.
The mass of particles that have been accelerated is  given by 
\begin{equation}
m_{AP} = \sum N_i A_i m_P 
\end{equation}
where $N_i$ is the total number of nuclei of species $i$, $A_i$ the corresponding mass number and $m_P$ the proton mass. The number $N_i$, or rather the product $N_i A_i$ can be determined by noticing that
the total energy carried by those accelerated particles is a fraction $f$ of the kinetic energy $E_0$ of the supernova 
 \begin{equation}
  \sum N_i A_i \int_0^{\infty} E \ Q(E) \ dE \ = \ f \ E_{0}
 \end{equation}
 where $Q(E)$ is the spectrum of accelerated particles ($A_i$ appears on the left side of Eq. 15 because
  energies are expressed in energy units {\it per nucleon}). The efficiency $f$ of conversion of $E_0$ to 
  accelerated particles {\it escaping the supernova} is estimated to $f\sim$0.1, while another
 0.1-0.2 goes to the acceleration of particles which are finally trapped in the SN (Ellison and Bykov 2011).
  The particle spectrum is often described by a power law in momentum 
   
\begin{equation}
Q(E) \ \propto \ \frac{p^{-s}}{\beta} {\rm exp}(-E/E_0)
\end{equation}
where $\beta=v/c$ is the velocity expressed as a 
fraction of the light velocity,
$p$ is the particle momentum per nucleon, 
the factor $s$ is usually 2$<s<$3 (in the case
of strong shocks) and $E_0$ is a cut-off energy, here taken to be 3 TeV (the results are insensitive
to much higher values).
  
  Assuming that the spectrum of accelerated particles escaping the SN is given by Eq. 16, one finds  
  the efficiency with which particles are accelerated from
  the shocked circumstellar medium, through Eq. 14 to 16:
 \begin{equation}
 W = \frac{m_{AP}}{M_{ACC}}
 \end{equation}
 and for rotating stars it is found to be in the range 3-6 10$^{-6}$ i.e. a few particles out of a million encountered ones are accelerated by the forward shock to GCR energies. In the case of non-rotating stars, the energetics is the same, but the swept-up mass is smaller (by a factor of two, on average, in order to get the observed \neo \ ratio) and the corresponding efficiency is $W\sim$10$^{-5}$. These estimates constitute
only a gross average, since the efficiency of particle acceleration may depend on several factors, not considered here,
like e.g.  the density of the circumstellar medium or the shock velocity - through a smoothly varying function
$f(v_S)$  instead of the Heavyside function $f(>\upsilon_{min})$=1 and $f(<\upsilon_{min})$=0  considered here - 
or the shock radius at the time of acceleration, since particles may subsequently suffer adiabatic cooling before
escaping (e.g. Drury 2011, Bykov and Ellison 2011). 
Notice that some of  those effects may have opposite time dependencies.  For instance,
particles accelerated earlier on (at higher shock velocities and presumably with higher efficiencies)  are expected
to suffer more from adiabatic cooling (because they are produced at smaller radii) . These effects require a much more
thorough investigation. {As a first step towards that direction, we also tested the case  where the efficiency of shock acceleration varies
with shock velocity as  $f \propto \upsilon^2$ (Drury 2011). In that case,  material with high   \neo \ is efficiently accelerated in the inner layers
of the most massive stars. In order to obtain again the observed GCR source \neo \ ratio, one has to dilute the mixture by allowing acceleration for shock
velocities lower then the reference values found above: we thus find values of $\upsilon_{min}$=1600 km/s for the HMM05 yields and 2150 km/s
for the CL06 yields; the corresponding overall acceleration efficiencies increase then by $\sim$40\% from the reference values given above,
ranging from 4-7 10$^{-6}$ for HMM05 yields to 1-1.4 10$^{-5}$ for CL06 yields.
The reason for obtaining such a small difference (only a few hundreds of km/s) between the non-realistic reference case and the -
perhaps more realistic - case of velocity dependent efficiency, is the adopted unified treatment, which considers acceleration
in both low mass and high mass supernova: the former accelerate almost pure ISM with low \neo \ and the latter almost pure wind material
with high \neo. Allowing for lower values of $\upsilon_{min}$ involves a considerably larger amount of ISM processed by 
low mass supernovae, since the lower acceleration efficiency ($f \propto \upsilon^2$) is more than compensated  by the  $n_{ISM} 4 \pi r^ 2 dr$
factor.

As approximate as they may be, the results obtained here  through the constrain of the observed GCR source \neo \ ratio,  
indicate clearly that acceleration has to occur for shock velocities larger than $\sim$1500 km/s, and they
may help to improve our understanding of particle acceleration in SN remnants.

\section{GCR cannot be accelerated (mainly) in superbubbles}

The idea that CGR are accelerated mainly in superbubbles has been suggested by Kafatos et al. (1981) and
reassessed by  Higdon et al. (1998).
Massive stars are mainly formed in OB associations and the winds of the most massive of them
initiate the formation of superbubbles, while the subsequent SN explosions power the expansion
of those superbubbles for a few 10$^7$ years. Higdon et al. (1998) argued that 
the environment of such superbubbles, enriched with the ejecta of stellar winds and core collapse
SN explosions, provides a composition that compares favourably to the inferred GCR source composition.

One potential problem with that idea is that $^{59}$Ni, a well known product of SN nucleosynthesis,
is absent from GCR arriving on Earth (Wiedenbeck et al. 1999). Since $^{59}$Ni is unstable to electron capure, with a lifetime of 
$\sim$10$^5$ y, Higdon et al. (1998) argued that SN explosions occur within the superbubble with a 
sufficiently low frequency (less than one SN every 3 10$^5$ y), as to allow for the decay of $^{59}$Ni
between two SN explosions. Prantzos (2005) pointed out that energetic stellar winds can also accelerate particles
and they are not intermittent (like SN) but occur continuously in superbubbles; in that case $^{59}$Ni is 
continuously  accelerated to GCR energies and, being unable to capture an electron, it becomes effectively stable
and it should be detectable in GCR. Binns et al. (2008) counter-argued that the period of energetic WR winds
represents a small (albeit not negligible) fraction in the early lifetime of an OB association and therefore only
little (and presumably undetectable) amounts of $^{59}$Ni would be accelerated, thus saving the "superbubble paradigm".

In the meantime, Higdon and Lingenfelter (2003) evaluated the isotopic composition of \neo \ - the critical abundance anomaly
in GCR -  expected in a superbubble, based on (a very heterogeneous set of)
then available yields of WR stars and SN. They found that the GCR \neo \ ratio
"...can be easily understood as the result of GCR accelerated primarily out of superbubbles with a mean metallicity
Z$_{SB}$=2.7$\pm$0.4 \zs..." and that this result "... provides strong, additional evidence for a superbubble origin of GCR".
In a subsequent paper, Linfenfelter and Higdon (2007) explored quantitatively that senario for other GCR abundances, finding
again that results compare favourably to observations. 

Despite their apparent sophistication, the aforementioned arguments for a superbubble origin of GCR miss a simple point:
{\it massive stars are the principal source of both \nea \ and \neb \ in the Universe}. As a result, the \neo \ ratio of a generation
of massive stars (integrated over an IMF and including WR winds and all kinds of SN ejecta) should be solar\footnote{Strictly speaking,
this concerns massive stars of roughly solar metallicity, such as those that contributed to the composition of the solar system; however,
metallicity has evolved very little in the past several billion years in the local volume of $\sim$ 1-2 kpc radius, 
where most GCR originate.}. Since  \neb \ is overabundant with respect to \nea \ only in the massive star winds 
(and not in the SN ejecta or the ISM),  a popular  exercice in the past 30 years  or so consisted in evaluating the mixing ratio
of WR ejecta with SN ejecta (or with average ISM), as to obtain the observed \neo \ ratio in GCR. 
(see Introduction and references therein). 

However,  it is expected that a properly
weighted mixture of the ejecta of massive stars in a superbubble (i.e. including stellar winds and SN ejecta and folded with a
stellar IMF) would produce a solar \neo \ ratio. In this section, the same exercise is repeated with the yields
of CL05 and HMM05 and by taking into account the corresponding stellar lifetimes. In Fig. \ref{Fig:Ne_yields}
we present again the adopted \nea \ and \neb \ yields of LC06 and HMM05  for the stellar winds (solid and dotted curves, i.e. the same
as in Fig. 3) and also the total  yields  (winds plus SN ejecta, points). The latter are from LC06 for stars with mass loss (but no rotation)
and from Woosley and Weaver (1995, WW95) for stars without mass loss; they are in good overall agreement, except for \nea \ in stars
around 20 \ms.  For the SN ejecta, we consider only stars with M$<$40 \ms,
i.e. we assume that more massive stars eject their \nea \ and \neb \ through their winds and then form
black holes; in that case we maximize the \neo \ ratio expected from massive stars.

  \begin{figure}
   \centering
 \includegraphics[width=0.49\textwidth]{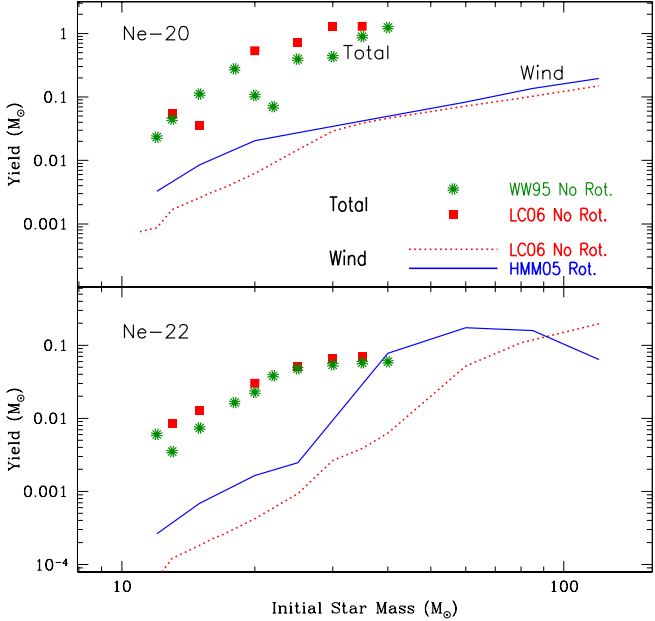}
 \caption{Yields   of \nea \ ({\it top}) and \neb \ {\it bottom})   as a function of the
 initial stellar mass.  Wind yields  are  from  HMM05 ({\it solid curves}) and LC06 ({\it dotted curves}), respectively
 (same as in Fig. 3). Total yields are from LC06 ({\it filled squares}) and WW95 ({\it asterisks}). 
 }
         \label{Fig:Ne_yields}
   \end{figure}

An inspection of Fig. \ref{Fig:Ne_yields} shows that the total (SN) \neb \ yields of M$<$40 \ms \ stars are
comparable to the wind \neb \ yields of the most massive ($\sim$100 \ms) stars; however, the corresponding
total \nea \ yields of M$<$40 \ms \ stars are
at least 10 times larger than the wind \nea \ yields of the most massive ($\sim$100 \ms) stars.
Thus, while the production of \neb \ receives a sizeable contribution from  the most massive 
stellar winds (at least for the rotating stars),
the production of \nea \ is totally dominated by the SN ejecta of M$<$40 \ms \ stars. In a coeval stellar
population, like the one expected in an OB association or a superbubble, the \neo \ ratio 
will evolve then from higher than solar to $\sim$solar values, 
as the M$<$40 \ms \ stars eject their core products a few Myr after
their more massive counterparts.

  \begin{figure}
   \centering
 \includegraphics[width=0.49\textwidth]{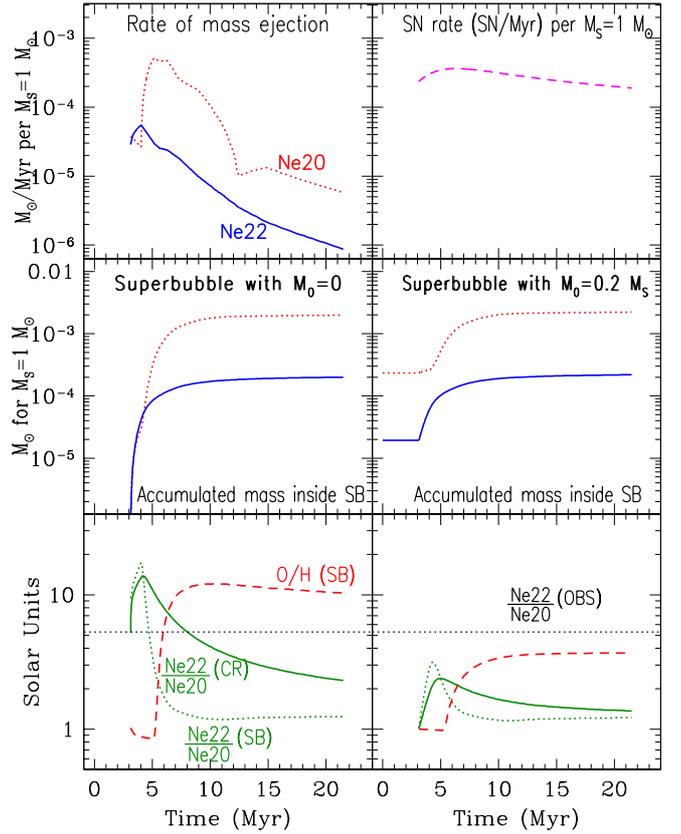}
 \caption{Evolution of composition in a superbubble. {\it Top left:} Ejection rate
 of \nea ({\it dotted}) \ and \neb \ ({\it solid}) in \ms /Myr per 1 \ms \ of stars formed (assuming a Salpeter IMF between 0.1 and 120 \ms).
 {\it Top right:} Supernova rate per Myr.
 {\it Middle}: Total mass of \nea \ ({\it dotted})  and \neb \ ({\it solid}) accumulated in the superbubble, in case
 there was no initial gas there (M$_0$=0, {\it left}) and in the presence of an initial mass of gas equal to 1/5 of the
 formed stars (M$_0$=0.20,{\it right}); in the former case \neb \ dominates by mass early on, whereas in the latter the mass of \nea \ dominates
 from the very beginning. {\it Bottom}: Corresponding evolution of the O/H ({\it dashed}) and of the (\neo)$_{SB}$  ratio  ({\it dotted}) in the SB gas;  the
 ({\it solid} curve represents the (\neo )$_{CR}$ ratio of particles accelerated {\it up to} that time and has to be compared to the observed GCR source
 ratio ({\it dashed horizontal line}). 
 }
         \label{Fig:SBevol}
   \end{figure}

We quantitative illustrate this effect in Fig. \ref{Fig:SBevol}, where we present the evolution of
a stellar population born at time $t$=0 with a Salpeter IMF and total mass of 1 \ms \ (results can be directly
scaled to masses of  OB associations, while abundances and abundance ratios remain the same).
The top right panel displays the rate of SN explosions; for a total stellar
mass of 10$^4$ \ms, resulting in a $\sim$100 OB stars, there are about 3 SN/Myr, i.e. $\sim$1 SN every 300 000 yr, as
evaluated in Higdon et al. (1998) in order to avoid the problem of $^{59}$Ni acceleration.
The top left panel displays the ejection rates of \nea \ and \neb \ (in \ms/Myr per \ms \ of stars formed) : \neb \ 
from stellar winds dominates only for a couple of Myr but  $\sim$5 Myr after the stellar formation, \nea \ from
M$<$40 \ms explosions dominates.   Notice that, in order to maximise the \neo \ ratio, we adopted only the
yields of HMM05 for these calculations (i.e. corresponding to the solid curves in Fig. \ref{Fig:Ne_yields} with
only wind yields above M=40 \ms).

The instantaneous (\neo)$_{SB}$ \ ratio in the superbubble at time $t$ is the ratio of the amounts of \nea \ and \neb \ cumulated up to that time,
which are provided in the middle panels of Fig. \ref{Fig:SBevol}.
In order to evaluate those quantities, one needs to make an assumption about the amount
of pre-existing gas still left inside the superbubble. In the left middle panel of Fig. \ref{Fig:SBevol}
it is assumed that no gas is left after star formation: gas is exclusively supplied by the wind and explosive stellar ejecta.
In that case,  \neb \ dominates early on, but \nea \ takes over after a couple of  Myr. In the right middle panel it is assumed that an amount of gas with solar composition
and equal to 20\% of the mass of formed stars  remained in the superbubble; the stellar ejecta are diluted into it and the superbubble
composition is now dominated always by \nea.

The bottom panels of Fig. \ref{Fig:SBevol} display the corresponding (\neo)$_{\rm SB}$ \ ratios in the superbubble. 
In the case of zero initial gas (left), the instantaneous  (\neo)$_{\rm SB}$ \ ratio  remains high - 
from 18 to 5 times solar -  for a couple of Myr, but soon after stars of M$<$40 \ms \ start dying its value goes rapidly to solar
(as expected from the nucleosynthesis argument presented in the beginning of this section).
The (\neo)$_{\rm CR}$ \ ($<t$) ratio of particles  accelerated to CR energies {\it up to time t}  is the time integral 
of the instantaneous (\neo)$_{\rm SB}$ \ ratio in the superbubble
\begin{equation}
\left(\frac{^{22}{\rm Ne}}{^{20}{\rm Ne}}\right)_{\rm CR}(<t) \ = \ \frac{1}{t} \int_0^t  \left(\frac{^{22}{\rm Ne}}{^{20}{\rm Ne}}\right)_{\rm SB} dt'
\end{equation}
where it is assumed that the efficiency of particle acceleration does not vary with time. 
 (\neo)$_{\rm CR}$ \ ($<t$) represents the mean CR composition accelerated by a superbubble which "operates" up to time $t$.

In the bottom left panel it is seen that the  (\neo)$_{\rm CR}$ \ ($<t$) ratio (in solar units) remains above the observed $R_{Obs}$=5.3$\pm$0.3  for
about 7 Myr, while it declines steadily after that, tending to an asymptotic value of $\sim$2, considerably smaller than observed. 
The corresponding metallicity in the superbubble, expressed by the O/H ratio (dashed curve),  is quite high - about 10 times solar - 
since it represents pure core collapse SN ejecta; such high metallicity values have never been reported for any astrophysical environment
(except SN remnants). 

The bottom right panel of Fig. \ref{Fig:SBevol} displays the corresponding quantities in the more realistic case of a superbubble
endowed with some gas left over from the star formation. In that case, although metallicity  is still high (about 3 times solar
after 7 Myr), the  \neo \ ratios never get above 3 times solar, either in the superbubble or in the accelerated particles; in fact they
are close to  solar for the largest part of the superbubble lifetime.

The results displayed in Fig.  \ref{Fig:SBevol}  suggest that the observed $R_{Obs}$=5.3$\pm$0.3 ratio in GCR cannot be
obtained from material accelerated in a superbubble: even under the most favorable possible conditions, such as those
adopted here (yields from {\it rotating} massive stars, only wind yields considered above M=40 \ms, no gas left over from star
formation), the resulting \neo \ ratio remains at high values only in the early evolution  of the superbubble, during a small
fraction of its lifetime. It is interesting to notice in that respect that Binns et al. (2008) suggested that {\it no substantial particle acceleration must occur 
during that early period}, in order to avoid the problem of $^{59}$Ni acceleration  from the WR winds; thus, saving
the "superbubble paradigm" for the origin of cosmic rays, requires that acceleration occurs only in the late superbubble evolution,
when the WR winds have essentially stopped and only SN inject intermittently their kinetic energy in the superbubble. However,  we have shown here
that during this period, the (\neo)$_{CR}$ \ ratio is considerably lower than observed. In fact, under realistic conditions (some initial gas left,
explosive yields also above 40 \ms \ considered,  at least for some stars), the \neo \ ratio in a superbubble would never get to values as high
as observed, contrary to claims made in the literature in the past decade. It must be stressed that this conclusion
does not depend on detailed adopted yields, but on a simple argument, namely that the IMF averaged \neo \ ratio
in a superbubble has to be close to solar during the largest period of the superbubble lifetime.

\section{Summary}

In this work we explore some implications of the idea that cosmic rays
are accelerated by the forward shock in supernova remnants, during their ST phase.
We focus on the chemical composition resulting from such an acceleration and, in particular, 
on the \neo \ ratio, which is the most characteristic feature of the  observed GCR source composition and 
is unaffected by atomic effects. For that purpose, we adopt recent models of the nucleosynthesis and
evolution of massive stars with mass loss: those of LC06 with no rotation and those of HMM05 with rotation.

In Sec. 2 we present a detailed summary of the properties of those models and, in particular, of the
chemical composition of their winds, insisting on the fact that rotating models release more
\neb \ in their winds than non-rotating ones. We also present the adopted model for the evolution of a
SN remnant within a stellar wind, based on the ideas of Ptuskin and Zirakashvili (2005) and the equations summarized in Caprioli (2011).

In the framework of the simple model adopted here, we follow the time-dependent composition of  GCR,
accelerated by the forward shock as it runs through either the ISM (in the case
of stars with M$<$25-35 \ms, depending on rotation) or through the stellar wind (in the case of more massive stars).
In fact, during the largest part of the ST phase, the shock runs through ISM and encounters a $\sim$solar composition.
In order to reproduce the observed high value of (\neo)$_{CR}$    ($R_{Obs}$=5.3$\pm$0.3 in solar units)  after accounting for
the stellar IMF, we have then to assume that acceleration is efficient only during a short early period in the ST phase.
We chose to use the shock velocity as a criterion for efficient acceleration and, based on the aforementioned CR
composition argument, we find that shock velocities larger than $\sim$1900 km/s (for the rotating stellar models) or
2400 km/s ( for the non-rotating ones) are required. This result is obtained by 
assuming a step function for the efficiency $f$ of particle acceleration
($f$=0 before the ST phase and after $\upsilon_{min}$, and $f$=1 between the two).
 For the - perhaps, more realistic - 
 assumption of a velocity-dependent efficiency  $f\propto \upsilon^2$, we find slightly lower values for
 $\upsilon_{min}$ (1600 km/s for HMM05 yields and 2150 km/s for CL06 yields, respectively).
 In the framework of the adopted models, this corresponds to a circumstellar
mass of  several tens of \ms \ encountered by the forward shock.  
Assuming, furthermore, that 10\% of the SN kinetic energy is used in 
acceleration of escaping cosmic rays with standard energy spectra, allows us to evaluate the efficiency of that acceleration:
we find that a few particles out of a million encountered by the forward shock  are accelerated to CR energies.
We also notice that this scheme of GCR acceleration does not suffer from problems related to the absence of unstable
$^{59}$Ni in observed GCR composition: this heavy nucleus is well inside the SN ejecta and is not reached by the
forward shock which accelerates only wind material and ISM.

The aforementioned scenario assumes that even the most massive stars, up to 120 \ms, develop
  strong forward shocks and accelerate the particles of their WR winds. For non-rotating stars, this
  is a rather extreme assumption, since it has been argued (Heger et al. 2003) that non-rotating masssive
  stars of about solar metallicity collapse into black holes, if their
  mass is in the 30-60 \ms \ range (see their Fig. 1). Notice that stars in the 60-120 \ms \ range (the most
  important \neb \  producers) end as  black
  holes in their scheme. However,  for slightly higher metallicities - such as those
  encountered in the inner Galactic disk - they find that  only neutron stars are formed,
  because higher stellar mass losses result in a less massive star at explosion. It should be noticed that
  the details of massive star explosions remain poorly understood at present (see e.g. Hanke et al. 2011) and so is the fate of a massive star above 30 \ms \ (see Fryer et al. 2011 for a recent 
  - but certainly not  definitive - assessment). 
 The fate of the rotating mass losing stars considered here is even less well known.
  
  In view of the  aformentioned uncertainties, we feel that the scenario proposed here can be considered as valid
  at present, although future refinements in our understanding of massive star explosions may change it quantitatively
  (and even qualitatively, if it turns out that most masive stars above, say, 50 \ms, end up as black holes).

Finally, we explore the idea that CR are accelerated in superbubbles, in which case their composition results from the
ejecta of both stellar winds and SN explosions. We first notice that simple nucleosynthesis arguments suggest that
the resulting composition, averaged over the stellar IMF, should be very close to solar. We  demonstrate this quantitatively,
with a simple model for the evolving composition of a superbubble, enriched first by the (\neb \ rich) winds of the
most massive stars, then by the (\nea \ rich) SN ejecta of less massive stars. We find that, after a few Myr the superbubble (\neo)$_{SB}$  \ ratio
tends to solar, and so does the average (\neo)$_{CR}$ \ ratio in accelerated particles. We conclude that superbubbles cannot provide
the observed high  $R_{Obs}$ value of CR sources and, therefore, are not the main site of CR acceleration.
 On the contrary, SN remnants - including those expanding in the pre-explosion environment of a stellar wind - appear
 as suitable sites of GCR acceleration.

\begin{acknowledgements}

I am grateful to G. Meynet and M. Limongi for providing tables with yields and other properties 
of their stellar models and to the referee, L. Drury, for  constructive remarks. 

\end{acknowledgements}

{}


\end{document}